\documentclass[10pt,letterpaper,twocolumn]{article} 

\usepackage{ol2}
\usepackage{amsmath}
\usepackage{amssymb}
\usepackage{units}
\usepackage{graphicx}
\usepackage{bm}

\begin{document}

\twocolumn[

\title{Peak intensity measurement of relativistic lasers via nonlinear Thomson scattering}

\author{Omri Har-Shemesh and Antonino Di Piazza$^*$}

\affiliation{
Max-Planck-Institut f\"{u}r Kernphysik, Saupfercheckweg 1, 69117 Heidelberg, Germany\\
$^*$Corresponding author: dipiazza@mpi-hd.mpg.de
}

\begin{abstract}
The measurement of peak laser intensities exceeding $10^{20}\;\text{W/cm$^2$}$ is in general a very challenging task. We suggest a simple method to accurately
measure such high intensities up to about $10^{23}\,\text{W/cm$^2$}$, by
colliding a beam of ultrarelativistic electrons with the laser pulse. The method exploits
the high directionality of the radiation emitted by
ultrarelativistic electrons via nonlinear Thomson scattering. Initial electron
energies well within the reach of laser wake-field accelerators are required,
allowing in principle for an all-optical setup. Accuracies of the order of
10\% are theoretically envisaged.  
\end{abstract}


 ]

Since the successful application of the chirped pulse amplification technique to
generate short laser pulses \cite{Mourou1985}, records in terms of
peak intensity are being continuously set. Intensities as high as
$2\times 10^{22}\;\text{W/cm$^2$}$ have already been
reached~\cite{Hercules}, and much higher ones are
envisaged \cite{Trines_2010}, allowing for investigations in
different fields including strong-field quantum
electrodynamics (QED) and plasma physics \cite{Di_Piazza_2011}.

The analysis of experiments employing ultrarelativistic optical laser pulses, 
i.e. optical pulses with intensities exceeding $10^{20}\;\text{W/cm$^2$}$, requires the
precise knowledge of quantities characterizing the pulse itself such as its peak
intensity. However, ultrarelativistic peak intensity measurements are
especially difficult because the damage threshold of the equipment is usually
far exceeded by such strong laser pulses.

To the best of our knowledge, the peak intensity of ultrarelativistic 
pulses is currently determined by measuring energy, duration 
and spot-size of the pulse, the latter however at a lower intensity. 
Therefore, this method is prone to errors since
the spot-size can be affected by increasing the pulse intensity
\cite{Link,Bahk2004,Hercules}. It is desirable to find
precise methods in order to measure directly 
the peak intensity of the laser field. There are several suggestions for direct \textit{in situ} measurements of
peak intensities by employing multiply-charged ions. The general idea behind these methods
is to place ions at the laser focus, with charge number selected according to the
expected intensity. Then, by measuring either the photo-ion
momentum distribution~\cite{Smeenk}, or the ionization
fraction~\cite{KeitelIons,Link}, one can determine the laser peak intensity. 
An alternative method is based on the emission of electrons ``born'' via
ionization within the laser pulse~\cite{Gao}.

In this Letter we suggest a novel method to determine the peak intensity of an
ultrarelativistic laser pulse, by measuring the angular aperture
of the radiation spectrum emitted via nonlinear Thomson scattering
by an ultrarelativistic electron
beam colliding head-on with the pulse. The method is practically 
insensitive to the precise temporal
shape of the laser pulse and it allows for single-shot intensity measurements with 
theoretical accuracies 
in principle of the order of 10\%. Units with $c = \hbar = 1$ and a metric with signature $+---$
are employed throughout.

Classically it is known that an accelerated ultrarelativistic electron 
(mass $m$ and charge $e<0$) emits radiation within
a narrow cone around its instantaneous velocity, with aperture
of the order of $1/\gamma(t)$, with $\gamma(t)\gg 1$ being its
instantaneous Lorentz factor \cite{LL_CFT}. We exploit the high directionality of the
radiation emitted by an ultrarelativistic electron 
inside a strong laser pulse to relate
the angular aperture of the emitted spectrum
to the peak intensity of the driving pulse. In order to
obtain an analytical formula we assume that the 
laser field can be approximated as a plane wave.
Our numerical results will show that this approximation
gives accurate results also for a tightly focused Gaussian beam. 
For the sake of simplicity we also temporarily 
neglect effects of radiation reaction (RR), i.e. of the
action of the radiation emitted by the electron
on the motion of the electron itself \cite{LL_CFT}.

We consider a plane-wave field linearly polarized along
the $x$ direction, propagating along the positive $z$ direction
and we describe it via the vector potential $\bm{A}(\varphi)=-(E_0/\omega_0)f(\varphi)\hat{\bm{x}}$,
where $\varphi=\omega_0(t-z)$, $\omega_0$ is the central angular frequency of
the laser, $E_0$ its electric field amplitude and where $f(\varphi)$ is a shape function such 
that $f(0)=0$ and $|f(\varphi)|\le 1$. From the exact solution of the Lorentz
equation in such a plane-wave field \cite{LL_CFT} 
in the case of an ultrarelativistic electron with initial four-momentum
$p^{\mu}(\varphi=0)=p_0^{\mu}=\varepsilon_0(1,0,0,-\beta_0)$
with $\varepsilon_0=m\gamma_0$, $\beta_0=(1-1/\gamma_0^2)^{1/2}$ and $\gamma_0\gg 1$,
we can write the tangent of the angle $\theta(\varphi)$ between the instantaneous
velocity of the electron and the negative $z$-axis as
\begin{equation}
\label{eq:angle}
\tan\theta(\varphi) = -\frac{\xi_0}{\gamma_0}\frac{f(\varphi)}{1 - \xi_0^2f^2(\varphi)/4\gamma_0^2},
\end{equation}
where $\xi_0=|e|E_0/m\omega_0$. Since the angle $\theta(\varphi)$ practically coincides with the instantaneous 
emission angle in the ultrarelativistic limit, Eq. \eqref{eq:angle} shows that 
the maximal emission angle $\theta_m$ is reached when $|f(\phi)|=1$. By introducing 
the peak laser intensity $I_0=E_0^2/4\pi$, we obtain the following simple relation
between $I_0$ and $\theta_m$:
\begin{equation}
\label{eq:main_result}
I_0 [10^{20}\;\text{W/cm$^2$}] = 0.28\left[\frac{\omega_0[\unit{eV}] \varepsilon_0[\unit{MeV}]\sin\theta_{\mathrm{m}}}{1+\cos\theta_\mathrm{m}}\right]^2,
\end{equation}
which is independent of the exact form of $f(\varphi)$.

\begin{figure}[htb]
\centerline{
\includegraphics[width=8.3cm]{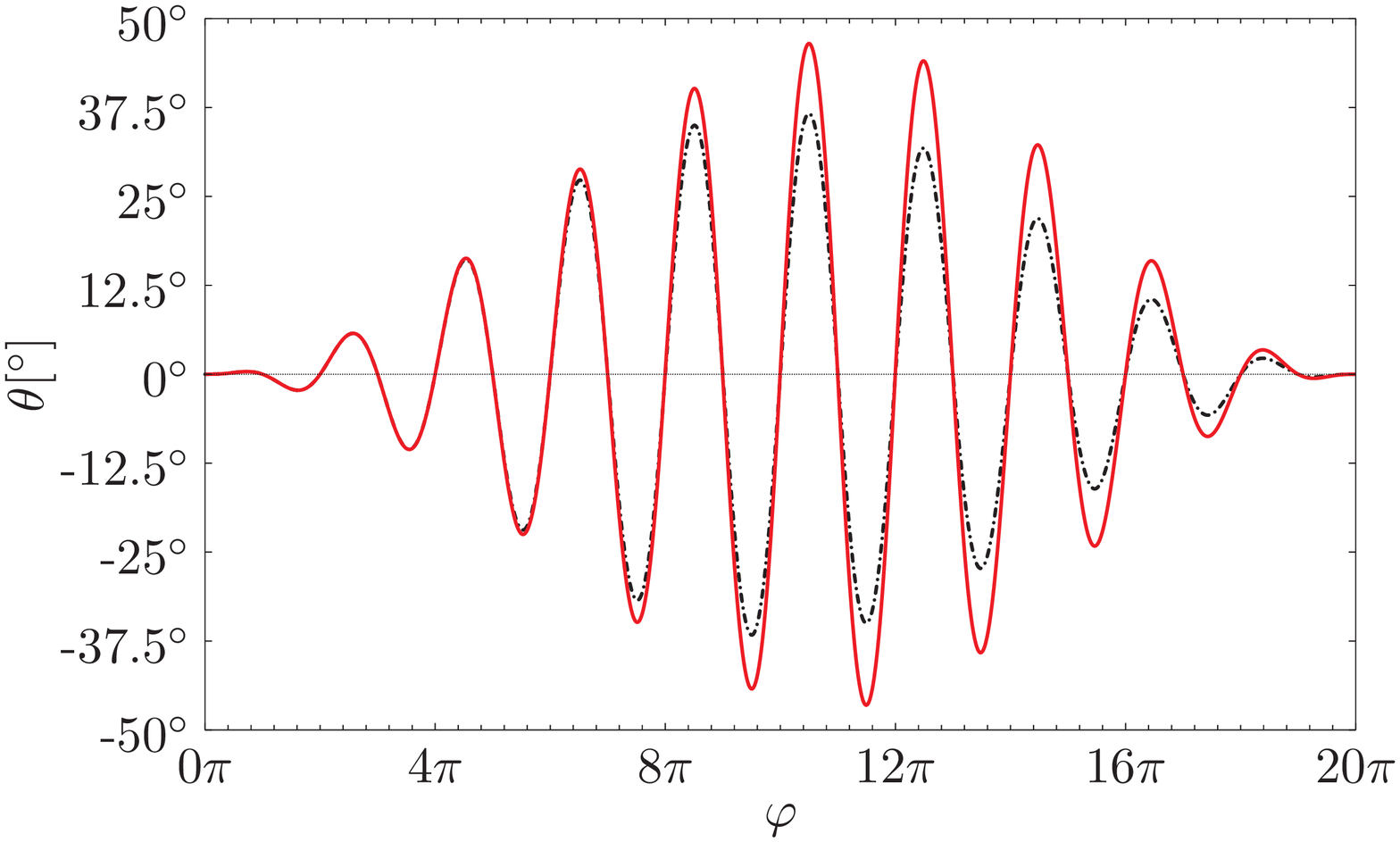}}
\caption{(Color online) Comparison of the direction of propagation of 
the electron with (solid red curve) and without (dash-dotted black curve) 
taking into account RR, as a function of the plane-wave 
phase $\varphi$. Numerical parameters are given in the text.} 
    \label{fig:LL_LF_Comparison} 
\end{figure}

As we have mentioned, we have so far neglected RR effects, which, in principle,
could become important at the intensities considered here \cite{Harvey_2011}. In the realm of classical 
electrodynamics such effects can be included by determining 
the electron trajectory via the so-called Landau-Lifshitz equation
\begin{equation}
\label{eq:LL_equation}
\begin{split}
\frac{du^\mu}{d\tau} &=  \frac{e}{m} F^{\mu\nu}u_\nu + \frac{2}{3}\frac{e^2}{m} \left[\vphantom{\frac{e^2}{m^2}} \frac{e}{m}(\partial_\alpha F^{\mu\nu})u^\alpha u_\nu - \right.\\
& - \left.\frac{e^2}{m^2} F^{\mu\nu}F_{\alpha\nu}u^\alpha + \frac{e^2}{m^2}(F^{\alpha\nu}u_\nu)(F_{\alpha\lambda}u^\lambda)u^\mu \right],
\end{split}
\end{equation}
where $u^{\mu}$ is the electron's four-velocity, $\tau$ its proper time, $F^{\mu\nu}(x)$ the background laser
field and $\alpha=e^2\approx 1/137$ the fine-structure constant \cite{LL_CFT,Di_Piazza_2011}. The Landau-Lifshitz equation has been solved exactly for a plane wave \cite{Antonino_exact_sol} and the solution in the above-described scenario provides
\begin{equation}
\label{eq:RR_angle}
\tan\theta(\varphi) = -\frac{\xi_0}{\gamma_0}\frac{\mathcal{I}(\varphi)}{1 -
[h^2(\varphi)-1+\xi_0^2\mathcal{I}^2(\varphi)]/4\gamma_0^2},
\end{equation}
where
\begin{align}
\label{eq:h_def}
h(\varphi) &= 1 + \frac{4}{3}\alpha\frac{\omega_0}{m}\gamma_0\xi_0^2\int_0^{\varphi}d\zeta [f'(\zeta)]^2,\\
\label{eq:I_def}
\mathcal{I}(\varphi) &= \int_0^{\varphi}d\zeta \left[ h(\zeta)f'(\zeta) +
\frac{4}{3}\alpha\frac{\omega_0}{m}\gamma_0f''(\zeta)\right],
\end{align}
with the prime denoting differentiation with
respect to its argument. By maximizing the expression of $\theta(\varphi)$ from Eq. (\ref{eq:RR_angle})
with respect to $\varphi$, one obtains the value of $\theta_m$, which includes RR effects. Equations (\ref{eq:RR_angle})-(\ref{eq:I_def}) show that: 1) RR
effects, being mainly a dissipative effect, will induce a dependence of
$\theta_m$ on the pulse-shape function $f(\varphi)$; 2) the expression of $\tan\theta(\varphi)$
in Eq. (\ref{eq:RR_angle}) reduces to that in Eq. (\ref{eq:angle}) in the formal limit
$\alpha\to 0$; 3) RR effects become important if $(4\pi/3)\alpha(\omega_0/m)\gamma_0\xi_0^2N\gtrsim 1$, with $N$ being the number of laser cycles in the pulse, i.e. if
$\varepsilon_0[\text{MeV}]I_0[10^{20}\;\text{W/cm$^2$}]N/\omega_0[\text{eV}]\gtrsim
1.5\times 10^5$. Since RR effects act mainly as a damping force,
which slows down the electron, we expect that their inclusion increases 
the predicted value of $\theta_m$ at a given $I_0$. This is confirmed by
the results in Fig. \ref{fig:LL_LF_Comparison}, which displays
the angle $\theta(\varphi)$ neglecting (dash-dotted black curve) and 
including (solid red curve) RR effects
for $f(\varphi)=\sin^2(\varphi/2N)\sin\varphi$ and for the following 
numerical parameters: $\omega_0=1.55\;\text{eV}$, $\xi_0=100$ ($I_0=4\times 10^{22}\;\text{W/cm$^2$}$), $\gamma_0=150$ ($\varepsilon_0=77\;\text{MeV}$) and $N=10$ 
(pulse duration of $27\;\text{fs}$). As expected, at the beginning of the motion 
the two curves coincide, but then larger values of $\theta(\varphi)$ are 
reached if RR effects are included.

Below, we present a numerical example showing the virtues of our method. We 
consider a Gaussian beam \cite{Salamin}, spatially focused at the
origin of the coordinate system, linearly polarized along
the $x$ direction, propagating along the positive $z$ direction and
with a $\sin^2$ time envelope initially centered at
$-N\lambda_0/2$, with $\lambda_0=2\pi/\omega_0$ being the laser central wavelength. 
The numerical parameters characterizing the beam are: 
waist size of $5\;\text{$\mu$m}$, $\omega_0 = 1.55\;\text{eV}$, 
$I_0=2\times 10^{22}\;\text{W/cm$^2$}$ ($\xi_0=69$) and $N=10$. Also, 
the electron is initially at the origin and
counterpropagates with respect to the laser beam
with an energy of $\varepsilon_0=23\;\text{MeV}$ ($\gamma_0=45$).
\begin{figure}[htb]
\centerline{
\includegraphics[width=8.3cm]{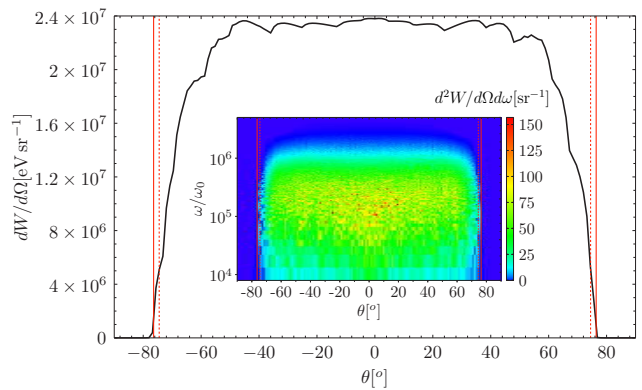}}
\caption{(Color online) Angular-resolved emitted energy spectrum $d^2W/d\Omega d\omega$ (inner
plot) and total energy emitted $dW/d\Omega$ per unit solid angle (outer plot), 
calculated for numerical parameters
given in the text. Vertical red lines indicate our theoretical
predictions for the maximal emission angle $\theta_m$, with (solid line) and 
without (dashed line) RR included.}
\label{fig:sim_result}
\end{figure}
The inner plot in Fig.~\ref{fig:sim_result} shows the single-electron emission spectrum \cite{Baier1998}
\begin{equation}
\label{Spectrum}
\frac{d^2W}{d\Omega d\omega}=\frac{e^2\omega^2}{4\pi^2}\bigg\vert\int dt\, \bm{n}\times(\bm{n}\times\bm{\beta}(t))e^{i\omega(t-\bm{n}\cdot\bm{r}(t))}\bigg\vert^2,
\end{equation}
i.e. the energy $W$ radiated per unit solid angle $\Omega$ 
and per unit frequency $\omega$, on the $x\text{-}z$ plane, while the outer
plot displays the corresponding integrated emitted energy $dW/d\Omega=
\int_0^{\infty}d\omega\,d^2W/d\Omega d\omega$. In Eq. (\ref{Spectrum})
$\bm{r}(t)$ and $\bm{\beta}(t)=d\bm{r}(t)/dt$ are the instantaneous 
position and velocity of the electron calculated 
via Eq. (\ref{eq:LL_equation}), and $\bm{n}$ is the observation direction. The two
vertical red lines indicate our theoretical expectations, one (dashed) without and
one (solid) with RR taken into account. The figure shows a
very good agreement between numerical and analytical results, although the
latter have been calculated in the plane-wave approximation.
Also, the inclusion of RR only increased the
predicted value of $\theta_m$ by about $2.2^\circ$. We note that the energy required for the
electron is well within the reach of presently available table-top laser-plasma
accelerators (LPAs) \cite{Esarey_2009}. The main advantages of using
a LPA are the microscopic dimensions of the electron bunches (transverse
radius and longitudinal length $\sim 10\;\text{$\mu$m}$ \cite{Esarey_2009}) permitting
an efficient overlapping with the laser beam and the allowance for a precise 
synchronization of the electron and the laser beams. The typical length
of the electron bunches generated via LPAs implies that 
the emission in the relevant part of the spectrum
(at energies $\gtrsim 10^4\;\text{eV}$, i.e. at wavelengths smaller 
than $0.1\;\text{nm}$, see Fig. \ref{fig:sim_result}) occurs incoherently. 
For this reason, in order to simulate an electron beam, we have performed simulations by 
varying the initial conditions of the electron. We have seen, for example, that by shifting the
initial transverse electron position within a radius of about 
$1\;\text{$\mu$m}$ from the origin, the angular extension of the
spectrum in Fig. \ref{fig:sim_result} remained practically unchanged,
while at larger radii it becomes narrower. Thus, by including conservatively
only those electrons in the bunch within a radius of $1\;\text{$\mu$m}$
and by assuming about $10^9$ electrons per bunch \cite{Esarey_2009},
we can estimate that in a single shot about $10^7$ would contribute to
the spectrum as shown in Fig. \ref{fig:sim_result}. By estimating the azimuthal angle
$\phi$ as $1/\gamma_0$, we conclude from the
numerical results in Fig. \ref{fig:sim_result} that about $10^5$
photons are expected to be emitted in a single shot at angles between, for example, $73^{\circ}$
and $75^{\circ}$, which are sufficient for a single-shot measurement
of the intensity. In this estimation we have used $\sim 10^5\;\text{eV}$
as average emitted photon energy.

By indicating as $\Delta F$ the uncertainty on the generic quantity $F$, Eq. (\ref{eq:main_result}) implies that 
\begin{equation}
\frac{\Delta I_0}{I_0}=2\sqrt{\left(\frac{\Delta \varepsilon_0}{\varepsilon_0}\right)^2+\left(\frac{\Delta \omega_0}{\omega_0}\right)^2+\left(\frac{\Delta \theta_m}{\sin\theta_m}\right)^2},
\end{equation}
showing that at $\gamma_0\approx \xi_0/2$ (so that $\sin\theta_m\approx 1$) the contribution of the uncertainty in $\theta_m$ is minimal. By safely neglecting $\Delta \omega_0/\omega_0$ and by assuming $\Delta \varepsilon_0/\varepsilon_0\approx 5\%$ \cite{Esarey_2009} and $\Delta\theta_m\approx 1^{\circ}$
(corresponding to the ratio between the dimension of a typical high-energy photon detector
$\sim 1\;\text{cm}$ and the distance $\sim 1\;\text{m}$ of the detector from the
interaction region), we obtain $\Delta I_0/I_0\approx 10\%$ in our numerical example. This theoretical estimate would in principle significantly improve the accuracy of intensity measurements, which nowadays are often even larger than $50\%$.

In the above analysis we have neglected quantum effects, which 
mainly amount in the ultrarelativistic regime to the photon recoil 
experienced by the electron in the photon emission \cite{Baier1998}.
These effects become important at $\chi=2\gamma_0E/E_{cr}\gtrsim 1$, 
with $E_{cr}=m^2/|e|=1.3\;\times 10^{16}\;\text{V/cm}$ 
being the ``critical'' field of QED \cite{Baier1998} and
in our numerical example it was $\chi=2\times 10^{-2}$.
In the convenient regime $\gamma_0\sim \xi_0/2$ quantum
effects can be neglected ($\chi<0.1$) for intensities smaller than 
$\sim 10^{23}\;\text{W/cm$^2$}$ at $\omega_0=1.55\;\text{eV}$, which determines the 
intensity upper limit of our method at least for an optimal accuracy.
In fact, at optical intensities larger than $10^{22}\;\text{W/cm$^2$}$
an electron emits on average more than one photon already in one laser period 
\cite{Baier1998}. The classical expression
of the emitted spectrum in Eq. (\ref{Spectrum}), which
is valid at $\chi\ll 1$, automatically takes into account multiple photon
emission \cite{Glauber_1951}. However, at $\chi\gtrsim 1$, i.e. when recoil
effects become important, there are not yet QED calculations on 
angular-resolved multiple-photon spectra in a strong laser field.

\end{document}